\documentclass[%
reprint,
superscriptaddress,
showpacs,
preprintnumbers,
graphicx,
floatfix,
byrevtex,
amsfonts,
amsmath,
amssymb,
aps,
prl,
]{revtex4-1}

\usepackage[]{graphicx}
\usepackage{epstopdf}
\usepackage{MnSymbol}
\usepackage{latexsym}
\usepackage{color}

\usepackage{dcolumn}
\usepackage{bm}
\usepackage[mathlines]{lineno}

\begin{document}

\title[Short Title]{Giant nonlinear optical effects induced by nitrogen-vacancy centers in diamond crystals}

\author{Mari Motojima}
\affiliation{Department of Applied Physics, Faculty of Pure and Applied Sciences, University of Tsukuba, 1-1-1 Tennodai, Tsukuba 305-8573, Japan}

\author{Takara Suzuki}
\affiliation{Department of Applied Physics, Faculty of Pure and Applied Sciences, University of Tsukuba, 1-1-1 Tennodai, Tsukuba 305-8573, Japan}

\author{Hidemi Shigekawa}
\affiliation{Department of Applied Physics, Faculty of Pure and Applied Sciences, University of Tsukuba, 1-1-1 Tennodai, Tsukuba 305-8573, Japan}

\author{Yuta Kainuma} 
\affiliation{School of Materials Science, Japan Advanced Institute of Science and Technology, Nomi, Ishikawa 923-1292, Japan}

\author{Toshu An}
\affiliation{School of Materials Science, Japan Advanced Institute of Science and Technology, Nomi, Ishikawa 923-1292, Japan}

\author{Muneaki Hase}
\email{mhase@bk.tsukuba.ac.jp}
\affiliation{Department of Applied Physics, Faculty of Pure and Applied Sciences, University of Tsukuba, 1-1-1 Tennodai, Tsukuba 305-8573, Japan}


\begin{abstract}
We investigate the effect of nitrogen-vacancy (NV) centers in single crystal diamond on nonlinear optical effects using 40 fs femtosecond laser pulses. The near infrared femtosecond pulses allow us to study purely nonlinear optical effects, such as optical Kerr effect (OKE) and two-photon absorption (TPA), relating to unique optical transitions by electronic structures with NV centers. It is found that both the nonlinear optical effects are enhanced by the introduction of NV centers in the N$^{+}$ dose levels of 2.0$\times$10$^{11}$ and 1.0$\times$10$^{12}$ N$^{+}$/cm$^{2}$. In particular, our data demonstrate that the OKE signal is strongly enhanced for the heavily implanted type-IIa diamond. We suggest that the strong enhancement of the OKE is possibly originated from cascading OKE, where the high-density NV centers effectively break the inversion symmetry near the surface region of diamond. 
\end{abstract}

\maketitle

\section{Introduction}
The optical and magnetic properties of the negatively charged nitrogen-vacancy (NV) center in diamond have been extensively investigated because of its potential application to quantum sensing \cite{Mizuochi:12}, such as magnetic and electric fields \cite{Pelliccione:16,Sasaki:16}, local temperature in bio-systems \cite{Clevenson:15}, and to quantum information technologies \cite{Kosaka:17}. In most cases, near-infrared luminescence upon the excitation by green laser can be controlled by the irradiation of microwave, which tunes the optical transition paths between the excited and ground states \cite{Naka:08}. To extend the optical property of diamond over the current generation of $linear$ optical regime, one requires ultrashort laser pulses, which enables us to induce coherent lattice vibrations \cite{Ishioka:06,Nakamura:16,Sasaki:18,Maehrlein:17} as well as nonlinear optical effect, such as the nonlinear refraction and nonlinear absorption \cite{Shen:01,Zhang:16,Almeida:17}. Almeida {\it et al} recently reported on nonlinear optical spectra in high-purity diamond using femtosecond laser pulses with photon energy from 0.83 to 4.77 eV and measured coefficients for the nonlinear refraction and nonlinear absorption \cite{Almeida:17}. Because of the absence of defect-related bands below the band-gap ($E_{g}$ = 5.5 eV), the two nonlinear optical effects were enhanced for the photon energy of $\hbar \omega$ = $E_{g}$/2. Although there are several investigations on nonlinear optical effects in bulk diamond and nano-diamond \cite{Dadap:91,Trojanek:10}, the effects of the NV center on the nonlinear optical phenomena have not yet been examined. However, this issue is important to further study new functionality of diamond photonics to advance nonlinear quantum sensing. 

Here we explore the NV center induced nonlinear optical effects in pure diamond, specifically optical Kerr effect (OKE) and two-photon absorption (TPA) in transparent region using 800 nm light, at which NV related bands will be sensitive. It is found that both the nonlinear optical effects are enhanced by the introduction of NV centers. 
Importantly, the signal enhancement is more visible for the time-resolved reflectivity measurement than that in the case of the conventional transmission measurements. 
We attribute the several times larger signal enhancement in the reflectivity mode observed at the highest dose level to dominant contribution from the cascaded OKE due to breaking the inversion symmetry near the surface region and minor one from TPA due to the NV center related bands in diamond. 

\section{Experimental}

We first carried out Z-scan measurement to examine nonlinear optical effects, such as OKE and TPA \cite{Bahae:90}. The Z-scan measurements were performed using a closed aperture mode, to study the nonlinear refraction and nonlinear absorption, respectively. 
As shown in Fig. 1(a), the transmittance of a diamond sample through an aperture (a diameter of $\approx$0.8 mm) in the far-field was measured as a function of the sample position $z$ with respect to the focal plane.
The light source used was a femtosecond regenerative amplifier system (RegA9040, Coherent), which produces $\approx$40 fs pulses at a central wavelength of $\lambda$$\approx$800 nm with an average power of $\geq$500 mW at 100 kHz repetition rate. The output of the amplifier was focused via a focal length $f$ = 50 mm lens into the samples. 
Z-scan measurement enables us to evaluate the nonlinear optical effects associated with the third-order nonlinear susceptibility $\chi^{(3)}$ from its transmittance change. 
The third-order nonlinear susceptibility can be expressed by: 
\begin{equation}
\chi^{(3)} = \chi^{(3)}_{bulk} + \chi^{(3)}_{NV} + \chi^{(2)}_{NV}\chi^{(2)}_{NV},
\end{equation}
where the first term $\chi^{(3)}_{bulk} $ describes bulk third-order susceptibility, the second term $\chi^{(3)}_{NV}$ represents NV-induced third-order susceptibility, and the last term suggests a possible contribution from the cascaded second-order susceptibility $\chi^{(2)}_{NV}$ \cite{Shen:01,Meredith:82,Mondal:18}, which may be detected by surface sensitive optical methods, such as optical reflectivity. Note that the second-order susceptibility $\chi^{(2)}$ is generally zero whenever the material has inversion symmetry, i.e., in the case for high purity diamond. The inversion symmetry can be, however, broken when NV centers are introduced, since the N atom and its adjacent vacancy break the inversion symmetry of the diamond crystal \cite{Maze:11}.
Eq. (1) is expected to be affected by both the OKE and TPA via Re$\chi^{(3)} = 2n_{0}^{2}\epsilon_{0}cn_{2}$ and Im$\chi^{(3)} = n_{0}^{2}\epsilon_{0}c\beta/k$, respectively, where $n_{2}$ is the nonlinear refraction, $n_{0}$ is the linear refraction coefficient, $\epsilon_{0}$ is the vacuum permittivity, $c$ is the speed of light in vacuum, $\beta$ represents nonlinear
absorption (TPA) coefficient, and $k$ denotes the wave vector \cite{Shen:01,Bahae:90}. 

\begin{figure}[htbp]
\centering\includegraphics[width=8.6cm]{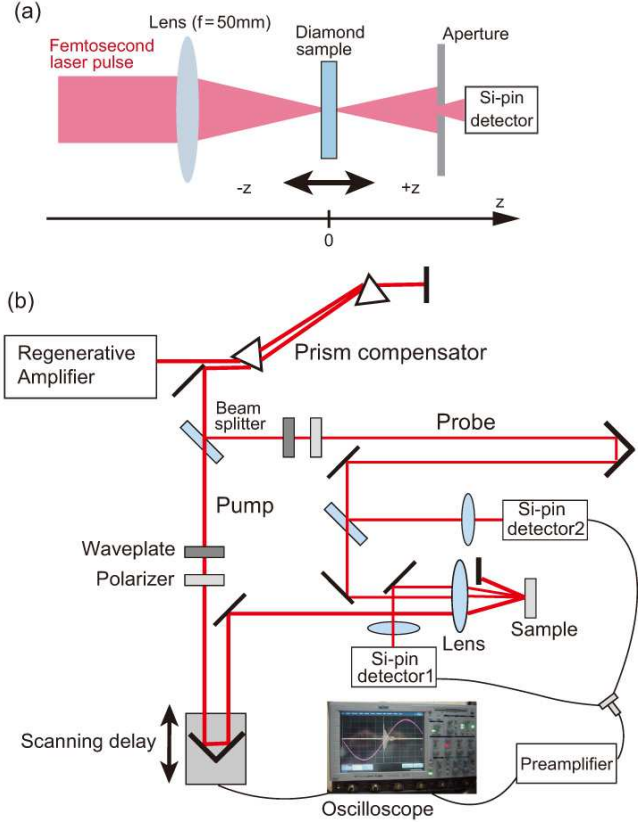}
\caption{
(a) The experimental set-up for Z-scan measurement using the closed aperture mode. (b) Schematics of femtosecond pump-probe experiment with the reflection mode. 
The difference of the photocurrent of the two Si-pin detectors was amplified, and averaged in a digital oscilloscope using the scanning time delay at 20 Hz. 
}
\end{figure}

To investigate the time-domain response from the OKE and TPA due to the NV centers near the sample surface, reflection-type pump-probe measurements for the diamond samples were also carried out using the same femtosecond regenerative amplifier system at room temperature by employing the fast scanning time-delay method [Fig. 1(b)] \cite{Hase:12}. 
The pump and probe beams were focused via the focal length $f$ = 100 mm lens onto the sample surface with a mutual angle of $\leq$5$^{\circ}$. The pump power was varied from 20 to 80 mW, while the probe power was kept at $\leq$ 2 mW. Polarization of the two beams was linear with orthogonal each other. 

The samples used were Element Six [100] type-IIa diamond crystal fabricated by CVD method, whose impurity (nitrogen: [N] and boron: [B]) levels were [N] $<$ 1 ppm and [B] $<$ 0.05 ppm, respectively. The sample size was 3.0 mm$\times$3.0 mm$\times$0.3 mm (thickness). To introduce NV centers 30 keV nitrogen ions (N$^{+}$) were implanted into the diamond samples at the N$^{+}$ dose levels of 2.0$\times$10$^{11}$ and 1.0$\times$10$^{12}$ ions/cm$^{2}$. The implanted depth deduced from the Monte Carlo calculation (TRIM) was about 30--40 nm \cite{Ziegler:85,Kikuchi:17} and the profile was close to a Gaussian function with FWHM of $\sim$ 50 nm. 
After that, the samples were annealed at 900--1000 $^{\circ}$C in an argon atmosphere for 1 hour to produce NV centers with a production efficiency of $\approx$ 1$\%$ (see Fig. 2) \cite{Pezzagna:10}. The production of NV centers was confirmed by optically-detected magnetic resonance (ODMR) technique using a 532 nm laser at $\approx$ 1 mW \cite{Kikuchi:17}. 

\begin{figure}[htbp]
\centering\includegraphics[width=8.8cm]{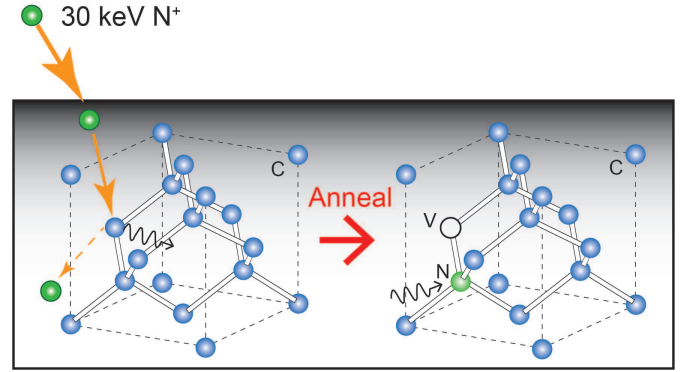}
\caption{Schematic for the production of NV centers by N$^{+}$ ion implantation and subsequent annealing in diamond crystalline samples. 
The N$^{+}$ ions implanted into diamond produce vacancies (V), which are required to make nitrogen-vacancy (NV) centers. Annealing the diamond sample induces diffusion of the vacancies, which can then be trapped by the implanted nitrogen atoms.}
\end{figure}

\section{Results and discussion}

Figure 3 shows the results of the closed-aperture Z-scan measurements obtained for different N$^{+}$ dose levels observed at the laser fluence of $I_{0}$$\approx$20 mJ/cm$^{2}$. 
Anti-symmetric line shape, whose peak and valley have not the same amplitude, was observed for all the samples.
The anti-symmetric lineshape, with a valley greater than the peak, indicates an existence of two-photon absorption \cite{Almeida:17}. 
The intensity of the transmittance change becomes larger when the NV centers were introduced at 2.0$\times$10$^{11}$ and 1.0$\times$10$^{12}$ N$^{+}$/cm$^{2}$.
In particular, the valley at $z$ = -- 0.2 mm exhibits $\approx$ 1.3 times larger transmission change compared to the pure diamond (non-implanted), indicating that absorption by NV centers plays a central role in the enhancement of the signals in Figs. 3(b) and 3(c). 

\begin{figure}[htbp]
\centering\includegraphics[width=8.6cm]{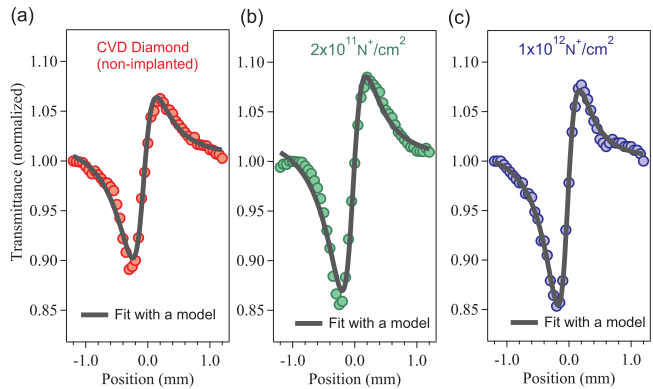}
\caption{Closed-aperture Z-scan results at $I_{0}$$\approx$20 mJ/cm$^{2}$ obtained for the pure diamond crystal and NV center introduced diamond at different dose levels, (a) non-implanted, (b) 2.0$\times$10$^{11}$N$^{+}$/cm$^{2}$, and (c) 1.0$\times$10$^{12}$ N$^{+}$/cm$^{2}$. The black solid lines represent the fit to the data using Eq. (2).
}
\end{figure}

We applied the general model of the Z-scan technique for the transmittance change \cite{Bahae:90,Yin:00},
\begin{equation}
T(x) = \\ 
1 +  \frac{4x}{( x^{2} + 9)(x^{2} +1)}\Delta \phi_{0} - \frac{2(x^2 + 3)}{( x^{2} + 9)(x^{2} +1)}\Delta\psi_{0},
\end{equation}
where $x=\frac{z}{z_{0}}$, $z$ is the position, $z_{0}=\pi w_{0}^{2}/\lambda$ is the Rayleigh length with $w_{0}$ being the beam waist, $\Delta \phi_{0} $ is induced phase shift and $\Delta\psi_{0}$ is a loss parameter. 
The parameters obtained by the fitting are listed in Table 1. 
The fitting parameter of $z_{0} \sim$ 0.2 mm corresponds to $w_{0} \sim$ 7 $\mu$m, a value being well consistent with that calculated ($\sim$ 8 $\mu$m) using $w_{0}$ = 0.61$\lambda$/$NA$, where $NA$ = 0.06 is the numerical aperture of the lens used. 
As seen in Table 1, the phase shift $\Delta \phi_{0} $ associated with nonlinear refraction $n_{2}$ becomes larger by a factor of $\approx$1.3 when the NV centers were introduced, while $\Delta\psi_{0}$ associated with nonlinear absorption $\beta$ (e.g., two-photon absorption) is enhanced by a factor of $\approx$1.6. 
Thus, we can see the small but significant increases in the nonlinear optical effects associated with the NV center on the Z-scan data, which corresponding to $\chi^{(3)}_{NV}$ in Eq. (1). 
To study further details, we used the relationships between the fitting parameters, $\Delta \phi_{0} $ and $\Delta\psi_{0}$, and nonlinear optical effects, which are $\Delta \phi_{0} $ = $kn_{2}I_{0}L_{eff}$ and $\Delta\psi_{0}$ = $\beta I_{0}L_{eff}/2$ \cite{Bahae:90}, 
where $L_{eff}$ is the effective length of the sample defined by $(1 - e^{-\alpha L})/\alpha$ with $\alpha$ ($\approx$0.1 cm$^{-1}$) \cite{E6} the linear absorption coefficient and $L$ (= 0.3 mm) the sample length. Using these relations, we obtain $n_{2}$ = 4.16$\times$10$^{-20}$m$^{2}$/W for the non-implanted diamond and $n_{2}$ = 5.50$\times$10$^{-20}$m$^{2}$/W for the NV center introduced diamond at 1.0$\times$10$^{12}$N$^{+}$/cm$^{2}$. 
In the same way, we obtain $\beta$ = 0.993$\times$10$^{-2}$cm/GW for the non-implanted diamond and $\beta$ = 1.61$\times$10$^{-2}$cm/GW for 
the NV center introduced diamond at 1.0$\times$10$^{12}$N$^{+}$/cm$^{2}$. These values are the same order of those obtained for high purity diamond \cite{Almeida:17}. 
Although the NV introduced samples exhibit the small enhancement of the nonlinear optical effects, the transmission mode might dominantly detect the bulk contribution $\chi^{(3)}_{bulk}$, and masked the precise signal from NV centers, $\chi^{(3)}_{NV}$ and $\chi^{(2)}_{NV}$, which are buried in the sample surface.  To examine the precise effect of $\chi^{(3)}_{NV}$ and $\chi^{(2)}_{NV}$, we show the results on the transient reflectivity measurements below.

\begin{table}
  \caption{The fitting parameters obtained using Eq. (2). The standard deviation of coefficients were obtained during the fitting procedure with Igor Pro.}
  \begin{tabular}{c|c|c|c}
  \hline
     Sample & $\Delta \phi_{0} $  & $\Delta\psi_{0}$ & $z_{0}$ (mm)\\
    \hline
     Diamond (non-implanted)  &  0.39$\pm$0.01 & 0.06$\pm$0.01 & 0.22$\pm$0.01 \\
        \hline
     Diamond (2.0$\times$10$^{11}$N$^{+}$/cm$^{2}$) & 0.53$\pm$0.02 & 0.08$\pm$0.01 & 0.22$\pm$0.01 \\
                 \hline
       Diamond (1.0$\times$10$^{12}$N$^{+}$/cm$^{2}$) & 0.52$\pm$0.01 & 0.10$\pm$0.01 & 0.20$\pm$0.01 \\
                 \hline
  \end{tabular}
\end{table}

Figure 4 shows the results of the time-resolved transient reflectivity ($\Delta R/R$) obtained by the pump-probe technique. 
The $\Delta R/R$ signal consists of an instantaneous drop, where the width of the negative peak (FWHM of $\approx$50 fs obtained by a Gaussian fit and deconvolution) nearly corresponds to the pulse duration of the laser used. 
The effectively smaller FWHM of the $\Delta R/R$ signal than that observed in a highly pure diamond \cite{Almeida:17} and that of the transient transmission ($\Delta T/T$) using UV excitation \cite{Liu:17} suggests the fact that the pulse duration in the present study is shorter than those in the literatures and the origin of the signal is due to nonlinear optical effects, such as OKE and TPA, which are governed by non-resonant intermediate states \cite{Shen:01}. 

\begin{figure}[htbp]
\centering
\includegraphics[width=8.6cm]{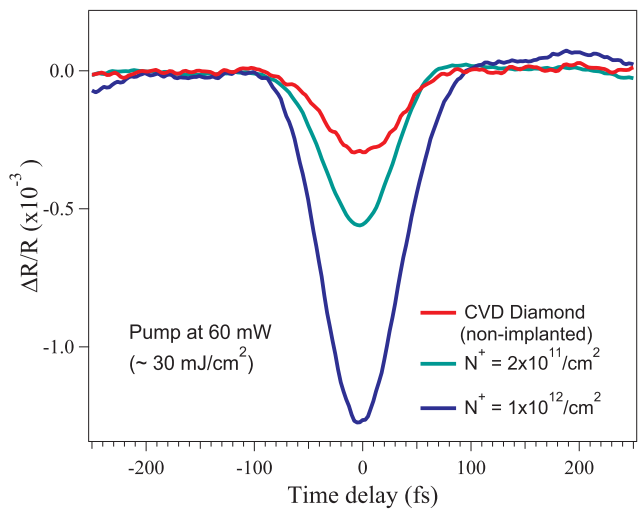}
\caption{Pump-probe reflectivity results obtained for the different dose levels of CVD diamond crystals at $I_{0}$$\approx$30 mJ/cm$^{2}$.}
\end{figure}

Although the transient negative peak signal is similar to the optical response reported by Nakamura {\it et al} \cite{Nakamura:16}, oscillations due to the generation of coherent optical phonons were not observed because of longer pulse width ($\approx$ 40 fs) than that ($\approx$ 10 fs) needed for the observation of the 40 THz coherent optical phonon \cite{Ishioka:06}.
Now the several times larger $\Delta R/R$ signal observed in the NV center introduced diamond indicates that the origin of the signal increase comes from the surface region rather than the bulk, since the signal enhancement was much smaller when we observed the transmittance by the Z-scan technique, in which we observe the nonlinear optical effects dominated by the native [N] and [B] defects in the bulk region. 
Note that we have also observed the $\Delta T/T$ signal using the pump-probe technique (data not shown). In this case, however, the factor of the signal enhancement was only less than several tens percent as in the case of the Z-scan measurements. Thus, the largest signal enhancement was observed only for the transient reflectivity signal, indicating that the signal enhancement is due to NV centers localized in the surface region. 

Under the assumption that the linear absorption (one-photon) for the near infrared light ($\lambda$$\approx$ 800 nm) can be neglected and the incident angle of the probe beam is the sample normal, the transient reflectivity change induced by the nonlinear optical effects can be expressed by the change in the real part of the refractive index $\Delta n$ \cite{Sabbah:02}, 
\begin{equation}
\frac{ \Delta R}{R} = \frac{4}{n_{0}^{2} - 1}\Delta n \approx 0.84(\Delta n_{_{OKE}} + \Delta n_{_{TPA}}),
\end{equation}
where $n_{0}$ = 2.4 for diamond at $\lambda$$\approx$ 800 nm was taken \cite{Zaitsev:01}. 
Here, $\Delta n$ is modulated by OKE, $\Delta n_{_{OKE}}$ = $n_{2}I_{0}$ \cite{Shen:01}, and 
free carrier contribution via TPA, $\Delta n_{_{TPA}}$, whose intensity dependence can be expressed as below based on the Drude model \cite{Sabbah:02},
\begin{equation}
\Delta n_{_{TPA}} = -\frac{2\pi e^{2}N}{n_{0}m^{*}\omega^{2}} = -\frac{2\pi e^{2}}{n_{0}m^{*}\omega^{2}} \frac{\beta I_{0}^{2}}{2\hbar\omega\tau_{p}} = \kappa\beta I_{0}^{2}. 
\end{equation}
where $e$ is the electron charge, $N$ is the free carrier density, $m^{*}$ is the electron effective mass, $\omega$ is the angular frequency of the laser used, $\tau_{p}$ is the pulse length, and $\kappa$ = -$2\pi e^{2}/(n_{0}m^{*}\omega^{2} \cdot 2\hbar\omega\tau_{p})$. 
As the pump fluence increases, nonlinear increase in the $|\Delta R/R|$ signal was observed for all the diamond samples, as shown in Fig. 5. 
By fitting the $|\Delta R/R|$ signals using a function of $aI_{0} + bI_{0}^{2}$, we obtained the fitting parameters $a$ and $b$ for different NV densities. Then, using the relationships from Eqs. (3) and (4), i.e. $a$ = 0.84$n_{2}$ and $b$ = 0.84$|\kappa|\beta$, the values of $n_{2}$ and $\beta$ are obtained as listed in Table 2. 
Although the value of $\beta$ shows small increase as the N$^{+}$ dose is increased, the value of $n_{2}$ associated with OKE increases more than an order of magnitude for the case of the heavily implanted diamond, 1.0$\times$10$^{12}$ N$^{+}$/cm$^{2}$. 
Note that the observed $n_{2}$ was negative in our study, which would be due to the contribution from cascading OKE as discussed in more details in the latter sections. 
Hereafter, we will discuss the value of $|n_{2}|$ to examine the effect of the NV center in diamond. The $|n_{2}|$ in Table 2 obtained for the non-implanted and the lightly implanted diamond (2.0$\times$10$^{11}$N$^{+}$/cm$^{2}$) are very small if we consider the standard deviations, but are in good agreement with those obtained by Z-scan measurements by other groups \cite{Kozak:12,Almeida:17}. 

\begin{figure}[htbp]
\centering
\includegraphics[width=8.6cm]{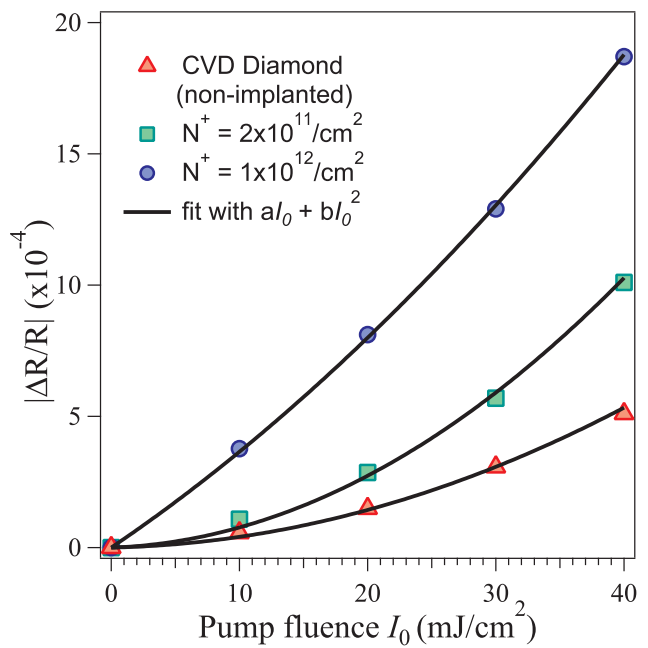}
\caption{Pump fluence dependence of the $|\Delta R/R|$ signal observed in diamond samples. The solid curves are the fit using a function of $aI_{0} + bI_{0}^{2}$. }
\end{figure}

\begin{table}
  \caption{The nonlinear refraction coefficient, $n_{2}$, and the nonlinear absorption coefficient, $\beta$, obtained for different N$^{+}$ dose levels, by using the fitting parameters described in the main text. The standard deviations of coefficients were obtained during the fitting procedure with Igor Pro.}
  \begin{tabular}{c|c|c}
  \hline
     Diamond sample & $|n_{2}|$ ($\times$10$^{-20}$m$^{2}$/W) & $\beta$ ($\times$10$^{-1}$cm/GW) \\
    \hline
     Non-implanted  & 0.73$\pm$0.63 & 0.90$\pm$0.06 \\
        \hline
     2.0$\times$10$^{11}$N$^{+}$/cm$^{2}$ & 1.20$\pm$0.90 & 1.75$\pm$0.09 \\
                 \hline
     1.0$\times$10$^{12}$N$^{+}$/cm$^{2}$ & 24.2$\pm$0.50 & 1.01$\pm$0.05 \\
                 \hline
  \end{tabular}
\end{table}

On the other hand, the quadratic dependence of the $|\Delta R/R|$ signal observed in the present study is surprisingly coincide with 
the signal from photo-carriers generated upon green laser (532 nm) illumination as the function of the green laser power \cite{Bourgeois:15}. 
Bourgeois {\it et al.} have detected two-photon excited free electrons in the conduction band (CB) as the photocurrent (photo-carriers)
both in type-Ib and -IIa CVD diamonds \cite{Bourgeois:15}. Since the quadratic dependence was enhanced for the pure type-IIa diamond, they concluded that the linear dependence dominated for the type-Ib diamond came from one-photon absorption via N$_{S}^{0}$ defect state \cite{Wotherspoon:03}. 
In our case, i.e., type-IIa diamond, however, the N$_{S}^{0}$ defects rarely exist and therefore the effect of N$_{S}^{0}$ defect states is negligible. Thus, the linear dependence on the $|\Delta R/R|$ signal should be dominated by the OKE. 
Thus, our data demonstrate that the OKE signal is enhanced much in the case of heavily implanted type-IIa diamond (1.0$\times$10$^{12}$N$^{+}$/cm$^{2}$). A possible reason for the giant enhancement of the OKE is the contribution from the cascading OKE \cite{Shen:01,Meredith:82,Mondal:18}, whose nonlinear susceptibility is expressed by $\chi^{(2)}_{NV}\chi^{(2)}_{NV}$ as in Eq. (1); the high density NV centers can effectively break the inversion symmetry near the sample surface region.  
Note that non-zero $\chi^{(2)}$ would generate second-harmonic generation (SHG) and therefore the incident pump photon energy cannot fully be supplied for the competing TPA \cite{DeSalvo:92}. 
In the case of the nonlinear refraction (e.g., cascading OKE), the coherence length $L_{coh}$ might be important, which is $\sim$16 $\mu$m, obtained using $L_{coh}=\pi/\Delta k$, where $\Delta k = k_{2\omega} - k_{\omega} $ is the wave vector mismatch during SHG \cite{DeSalvo:92}. In the case of the nonlinear absorption (e.g., TPA) via Im$\chi^{(3)}$, the Rayleigh length ($z_{0}$$\approx$200 $\mu$m in the present study) would play a major role. However, considering the signal enhancement occurred in the region where NV centers exist and inversion symmetry was broken ($\chi^{(2)}\neq 0$), our data indicate that the NV center introduced surface region with $L_{NV}$ $\sim$ 60--70 nm thick in maximum play a dominant role. 

\begin{figure}[htbp]
\centering
\includegraphics[width=8.7cm]{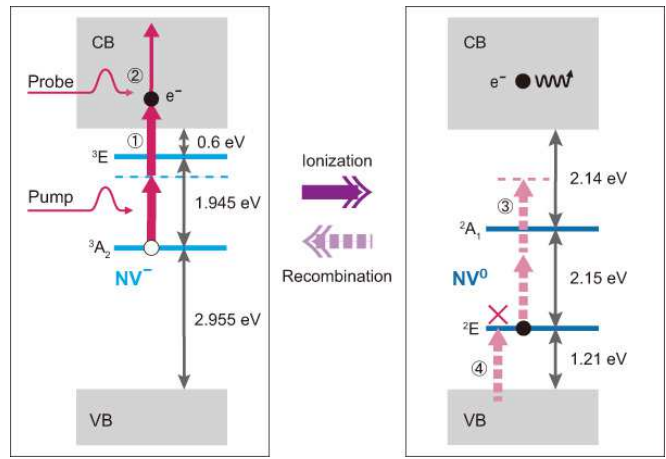}
\caption{Energy diagram for the NV$^-$ and NV$^0$ states in diamond. 
The NV introduced samples have natively the mid-gap electronic state of the negatively charged nitrogen-vacancy, i.e., NV$^-$ state (Left panel). The energy levels of the NV$^-$ and NV$^0$ states are defined by the binding energy of the NV center. The Fermi level of our sample in the NV$^-$ state (the nitrogen concentration of $\sim$10$^{17}$cm$^{-3}$) would be $\approx$4 eV above the valence band (VB) \cite{Collins:02}, and thereby electrons are populated at the $^{3}$A$_{2}$ ground state in the equilibrium. The lifetime of the $^{3}$E state is generally nanosecond time scale \cite{Fuchs:10}, however, in the present non-resonant case, intermediate state given by the dashed line should have very short lifetime (an order of the pulse length) and it does not significantly contribute to the TPA signal. 
The pumping action promotes an electron from the $^{3}$A$_{2}$ ground state into the conduction band (CB) via TPA (transition \textcircled{\scriptsize 1}), resulting in generation of a free electron, which is immediately followed by the formation of the neutrally charged NV$^0$ state (Ionization; right panel). Since femtosecond laser pulses can follow the ultrafast carrier dynamics in the CB, a part of the free carriers can absorb a probe photon (free carrier absorption; transition \textcircled{\scriptsize 2} ) when the pump and probe pulses overlap each other at $t$ = 0, resulting in the negative peak signal on $\Delta R/R$. For the NV$^0$ state, unlike in the case of irradiation of green laser, TPA (or one-phonon excitation) from the $^{2}$E ground state with a 1.55 eV probe photon (transition \textcircled{\scriptsize 3}) is hard to occur because of the energy mismatch and therefore recombination process via further excitation of electron from the VB into the $^{2}$E ground state (transition \textcircled{\scriptsize 4}) is not available. }
\end{figure}

By contrast, the nonlinear absorption, $\beta$, shows a largest value for the lightly implanted diamond (2.0$\times$10$^{11}$N$^{+}$/cm$^{2}$),   
indicating the stronger absorption of the probe light upon the photoexcitation, which can be discussed based on the band model of NV centers \cite{Collins:02,Fuchs:10}. 
Since the band-gap energy ($E_{g}$ = 5.5 eV) of diamond is much larger than the photon energy used (1.55 eV), and for high-purity diamond the two-photon absorption starts from $\approx$ $E_{g}$/2 \cite{Almeida:17}, it is required to consider the mid-gap states associated with the NV centers, which locates in between the band-gap, as illustrated in Fig. 6 \cite{Subedi:19}.  
For the negatively charged NV$^-$ state, the two-photon absorption (1.55 + 1.55 eV) would promote an electron from the $^{3}$A$_{2}$ ground state into the CB \cite{Aslam:13}. 
In this case, the two-photon absorption induces free carrier generation in the CB, and then immediately transfers the NV$^-$ to NV$^0$ states, i.e., the ionization \cite{Dhomkar:16}(see Fig. 6). 
Therefore, the existence of the NV centers will promote strong two-photon absorption as the pump action in the initial NV$^-$ state and subsequent transient one-photon absorption of a 1.55 eV probe photon by free carriers, which would induce a few times larger $\Delta R/R$ signal observed in NV center introduced diamond at 2.0$\times$10$^{11}$N$^{+}$/cm$^{2}$ (Fig. 4). 
Note that the ionization and/or recombination from the NV$^-$ to NV$^0$ states or vise versa have been discussed under the green laser (532 nm) irradiation \cite{Aslam:13,Dhomkar:18}. There, both the ionization (NV$^-$ $\rightarrow$ NV$^0$) and recombination (NV$^0$ $\rightarrow$ NV$^-$) rates exhibited quadratic dependence on the green laser power for relatively low densities of NV centers (type-IIa, [N] $<$ 1 ppm) \cite{Dhomkar:18}. The laser wavelength used in the present study is, however, near infrared (800 nm), 
and it is not possible to induce the recombination with such a low photon energy (see Fig. 6, right panel).
It is to be also noted that the transient $\Delta R/R$ signal observed in the present study is dynamical phenomenon, originating dominantly from the initial NV$^-$ state, and contribution from the NV$^0$ state to the $\Delta R/R$ signal will be negligibly small, although further experiments using different photon energies and theoretical analysis are required to fully understand the ultrafast nonlinear dynamics in diamond. 
In addition, for the heavily implanted diamond, the energy loss due to the SHG will be largest for the highest dose of 1.0$\times$10$^{12}$ N$^{+}$/cm$^{2}$, and thereby the probability of TPA would be decreased. 
This would explain why the $\beta$ value associated with the TPA decreases at 1.0$\times$10$^{12}$ N$^{+}$/cm$^{2}$ in Table 2. 

\medskip
\section{Conclusion}
We investigate the effect of nitrogen-vacancy (NV) centers in diamond on nonlinear optical effects using 40 fs femtosecond laser pulses. The near infrared femtosecond pulses allow us to study purely nonlinear optical effects, such as optical Kerr effect and two-photon absorption, relating to unique optical transitions by electronic structures involved with NV centers. It is found that both the nonlinear optical effects are enhanced by the introduction of NV centers in the N$^{+}$ dose levels of 2.0$\times$10$^{11}$ and 1.0$\times$10$^{12}$ N$^{+}$/cm$^{2}$. 
The signal enhancement was more visible for the time-resolved reflectivity measurement ($\Delta R/R$) than that in the case of the conventional transmission ($T$) measurements by the Z-scan technique. 
We attribute the several times larger signal enhancement in the $\Delta R/R$ mode observed at N$^{+}$ dose levels of 1.0$\times$10$^{12}$ N$^{+}$/cm$^{2}$ to dominant contribution from the cascading OKE originating from breaking the inversion symmetry near the surface region, and minor one from TPA due to the NV$^-$ bands in diamond. 
Our results will open the new window to develop all optical quantum sensing and computing based on nonlinear optical effects, such as a single-photon source using waveguide or nanopillar structures \cite{Babinec:10,Aharonovich:11}, and diamond ring resonators operating at telecom wavelengths \cite{Hausmann:14}, which are useful for quantum technologies in a wide range of spatial and frequency resolutions. 

\section*{Funding}
This work was supported by CREST, JST (Grant Number. JPMJCR1875), and JSPS KAKENHI (Grant Number. 17H06088), Japan. 

\section*{Acknowledgments}
The authors acknowledge Richarj Mondal of University of Tsukuba for his helpful comments on the Z-scan measurements. 

\bibliography{sample}

\begin{thebibliography}{41}%
\makeatletter
\providecommand \@ifxundefined [1]{%
 \@ifx{#1\undefined}
}%
\providecommand \@ifnum [1]{%
 \ifnum #1\expandafter \@firstoftwo
 \else \expandafter \@secondoftwo
 \fi
}%
\providecommand \@ifx [1]{%
 \ifx #1\expandafter \@firstoftwo
 \else \expandafter \@secondoftwo
 \fi
}%
\providecommand \natexlab [1]{#1}%
\providecommand \enquote  [1]{``#1''}%
\providecommand \bibnamefont  [1]{#1}%
\providecommand \bibfnamefont [1]{#1}%
\providecommand \citenamefont [1]{#1}%
\providecommand \href@noop [0]{\@secondoftwo}%
\providecommand \href [0]{\begingroup \@sanitize@url \@href}%
\providecommand \@href[1]{\@@startlink{#1}\@@href}%
\providecommand \@@href[1]{\endgroup#1\@@endlink}%
\providecommand \@sanitize@url [0]{\catcode `\\12\catcode `\$12\catcode
  `\&12\catcode `\#12\catcode `\^12\catcode `\_12\catcode `\%12\relax}%
\providecommand \@@startlink[1]{}%
\providecommand \@@endlink[0]{}%
\providecommand \url  [0]{\begingroup\@sanitize@url \@url }%
\providecommand \@url [1]{\endgroup\@href {#1}{\urlprefix }}%
\providecommand \urlprefix  [0]{URL }%
\providecommand \Eprint [0]{\href }%
\providecommand \doibase [0]{http://dx.doi.org/}%
\providecommand \selectlanguage [0]{\@gobble}%
\providecommand \bibinfo  [0]{\@secondoftwo}%
\providecommand \bibfield  [0]{\@secondoftwo}%
\providecommand \translation [1]{[#1]}%
\providecommand \BibitemOpen [0]{}%
\providecommand \bibitemStop [0]{}%
\providecommand \bibitemNoStop [0]{.\EOS\space}%
\providecommand \EOS [0]{\spacefactor3000\relax}%
\providecommand \BibitemShut  [1]{\csname bibitem#1\endcsname}%
\let\auto@bib@innerbib\@empty
\bibitem [{\citenamefont {Mizuochi}\ \emph {et~al.}(2012)\citenamefont
  {Mizuochi}, \citenamefont {Makino}, \citenamefont {Kato}, \citenamefont
  {Takeuchi}, \citenamefont {Ogura}, \citenamefont {Okushi}, \citenamefont
  {Nothaft}, \citenamefont {Neumann}, \citenamefont {Gali}, \citenamefont
  {Jelezko}, \citenamefont {Wrachtrup},\ and\ \citenamefont
  {Yamasaki}}]{Mizuochi:12}%
  \BibitemOpen
  \bibfield  {author} {\bibinfo {author} {\bibfnamefont {N.}~\bibnamefont
  {Mizuochi}}, \bibinfo {author} {\bibfnamefont {T.}~\bibnamefont {Makino}},
  \bibinfo {author} {\bibfnamefont {H.}~\bibnamefont {Kato}}, \bibinfo {author}
  {\bibfnamefont {D.}~\bibnamefont {Takeuchi}}, \bibinfo {author}
  {\bibfnamefont {M.}~\bibnamefont {Ogura}}, \bibinfo {author} {\bibfnamefont
  {H.}~\bibnamefont {Okushi}}, \bibinfo {author} {\bibfnamefont
  {M.}~\bibnamefont {Nothaft}}, \bibinfo {author} {\bibfnamefont
  {P.}~\bibnamefont {Neumann}}, \bibinfo {author} {\bibfnamefont
  {A.}~\bibnamefont {Gali}}, \bibinfo {author} {\bibfnamefont {F.}~\bibnamefont
  {Jelezko}}, \bibinfo {author} {\bibfnamefont {J.}~\bibnamefont {Wrachtrup}},
  \ and\ \bibinfo {author} {\bibfnamefont {S.}~\bibnamefont {Yamasaki}},\
  }\href {\doibase 10.1038/nphoton.2012.75} {\bibfield  {journal} {\bibinfo
  {journal} {Nature Photon.}\ }\textbf {\bibinfo {volume} {6}},\ \bibinfo
  {pages} {299} (\bibinfo {year} {2012})}\BibitemShut {NoStop}%
\bibitem [{\citenamefont {Pelliccione}\ \emph {et~al.}(2016)\citenamefont
  {Pelliccione}, \citenamefont {Jenkins}, \citenamefont {Ovartchaiyapong},
  \citenamefont {Reetz}, \citenamefont {Emmanouilidou}, \citenamefont {Ni},\
  and\ \citenamefont {Jayich}}]{Pelliccione:16}%
  \BibitemOpen
  \bibfield  {author} {\bibinfo {author} {\bibfnamefont {M.}~\bibnamefont
  {Pelliccione}}, \bibinfo {author} {\bibfnamefont {A.}~\bibnamefont
  {Jenkins}}, \bibinfo {author} {\bibfnamefont {P.}~\bibnamefont
  {Ovartchaiyapong}}, \bibinfo {author} {\bibfnamefont {C.}~\bibnamefont
  {Reetz}}, \bibinfo {author} {\bibfnamefont {E.}~\bibnamefont
  {Emmanouilidou}}, \bibinfo {author} {\bibfnamefont {N.}~\bibnamefont {Ni}}, \
  and\ \bibinfo {author} {\bibfnamefont {A.~C.~B.}\ \bibnamefont {Jayich}},\
  }\href {\doibase 10.1038/nnano.2016.68} {\bibfield  {journal} {\bibinfo
  {journal} {Nature Nanotech.}\ }\textbf {\bibinfo {volume} {11}},\ \bibinfo
  {pages} {700} (\bibinfo {year} {2016})}\BibitemShut {NoStop}%
\bibitem [{\citenamefont {Sasaki}\ \emph {et~al.}(2016)\citenamefont {Sasaki},
  \citenamefont {Monnai}, \citenamefont {Saijo}, \citenamefont {Fujita},
  \citenamefont {Watanabe}, \citenamefont {Ishi-Hayase}, \citenamefont {Itoh},\
  and\ \citenamefont {Abe}}]{Sasaki:16}%
  \BibitemOpen
  \bibfield  {author} {\bibinfo {author} {\bibfnamefont {K.}~\bibnamefont
  {Sasaki}}, \bibinfo {author} {\bibfnamefont {Y.}~\bibnamefont {Monnai}},
  \bibinfo {author} {\bibfnamefont {S.}~\bibnamefont {Saijo}}, \bibinfo
  {author} {\bibfnamefont {R.}~\bibnamefont {Fujita}}, \bibinfo {author}
  {\bibfnamefont {H.}~\bibnamefont {Watanabe}}, \bibinfo {author}
  {\bibfnamefont {J.}~\bibnamefont {Ishi-Hayase}}, \bibinfo {author}
  {\bibfnamefont {K.~M.}\ \bibnamefont {Itoh}}, \ and\ \bibinfo {author}
  {\bibfnamefont {E.}~\bibnamefont {Abe}},\ }\href {\doibase 10.1063/1.4952418}
  {\bibfield  {journal} {\bibinfo  {journal} {Rev. Sci. Instrum.}\ }\textbf
  {\bibinfo {volume} {87}},\ \bibinfo {pages} {053904} (\bibinfo {year}
  {2016})}\BibitemShut {NoStop}%
\bibitem [{\citenamefont {Clevenson}\ \emph {et~al.}(2015)\citenamefont
  {Clevenson}, \citenamefont {Trusheim}, \citenamefont {Teale}, \citenamefont
  {Schr\"{o}der}, \citenamefont {Braje},\ and\ \citenamefont
  {Englund}}]{Clevenson:15}%
  \BibitemOpen
  \bibfield  {author} {\bibinfo {author} {\bibfnamefont {H.}~\bibnamefont
  {Clevenson}}, \bibinfo {author} {\bibfnamefont {M.~E.}\ \bibnamefont
  {Trusheim}}, \bibinfo {author} {\bibfnamefont {C.}~\bibnamefont {Teale}},
  \bibinfo {author} {\bibfnamefont {T.}~\bibnamefont {Schr\"{o}der}}, \bibinfo
  {author} {\bibfnamefont {D.}~\bibnamefont {Braje}}, \ and\ \bibinfo {author}
  {\bibfnamefont {D.}~\bibnamefont {Englund}},\ }\href {\doibase
  10.1038/nnano.2016.68} {\bibfield  {journal} {\bibinfo  {journal} {Nature
  Phys.}\ }\textbf {\bibinfo {volume} {11}},\ \bibinfo {pages} {393} (\bibinfo
  {year} {2015})}\BibitemShut {NoStop}%
\bibitem [{\citenamefont {Sekiguchi}\ \emph {et~al.}(2017)\citenamefont
  {Sekiguchi}, \citenamefont {Niikura}, \citenamefont {Kuroiwa}, \citenamefont
  {Kano},\ and\ \citenamefont {Kosaka}}]{Kosaka:17}%
  \BibitemOpen
  \bibfield  {author} {\bibinfo {author} {\bibfnamefont {Y.}~\bibnamefont
  {Sekiguchi}}, \bibinfo {author} {\bibfnamefont {N.}~\bibnamefont {Niikura}},
  \bibinfo {author} {\bibfnamefont {R.}~\bibnamefont {Kuroiwa}}, \bibinfo
  {author} {\bibfnamefont {H.}~\bibnamefont {Kano}}, \ and\ \bibinfo {author}
  {\bibfnamefont {H.}~\bibnamefont {Kosaka}},\ }\href {\doibase
  10.1038/nphoton.2017.40} {\bibfield  {journal} {\bibinfo  {journal} {Nature
  Photon.}\ }\textbf {\bibinfo {volume} {11}},\ \bibinfo {pages} {309}
  (\bibinfo {year} {2017})}\BibitemShut {NoStop}%
\bibitem [{\citenamefont {Naka}\ \emph {et~al.}(2008)\citenamefont {Naka},
  \citenamefont {Kitamura}, \citenamefont {Omachi},\ and\ \citenamefont
  {Kuwata-Gonokami}}]{Naka:08}%
  \BibitemOpen
  \bibfield  {author} {\bibinfo {author} {\bibfnamefont {N.}~\bibnamefont
  {Naka}}, \bibinfo {author} {\bibfnamefont {T.}~\bibnamefont {Kitamura}},
  \bibinfo {author} {\bibfnamefont {J.}~\bibnamefont {Omachi}}, \ and\ \bibinfo
  {author} {\bibfnamefont {M.}~\bibnamefont {Kuwata-Gonokami}},\ }\href
  {\doibase 10.1002/pssb.200879877} {\bibfield  {journal} {\bibinfo  {journal}
  {phys. stat. sol. (b)}\ }\textbf {\bibinfo {volume} {245}},\ \bibinfo {pages}
  {2676} (\bibinfo {year} {2008})}\BibitemShut {NoStop}%
\bibitem [{\citenamefont {Ishioka}\ \emph {et~al.}(2006)\citenamefont
  {Ishioka}, \citenamefont {Hase}, \citenamefont {Kitajima},\ and\
  \citenamefont {Petek}}]{Ishioka:06}%
  \BibitemOpen
  \bibfield  {author} {\bibinfo {author} {\bibfnamefont {K.}~\bibnamefont
  {Ishioka}}, \bibinfo {author} {\bibfnamefont {M.}~\bibnamefont {Hase}},
  \bibinfo {author} {\bibfnamefont {M.}~\bibnamefont {Kitajima}}, \ and\
  \bibinfo {author} {\bibfnamefont {H.}~\bibnamefont {Petek}},\ }\href
  {\doibase 10.1063/1.2402231} {\bibfield  {journal} {\bibinfo  {journal}
  {Appl. Phys. Lett.}\ }\textbf {\bibinfo {volume} {89}},\ \bibinfo {pages}
  {231916} (\bibinfo {year} {2006})}\BibitemShut {NoStop}%
\bibitem [{\citenamefont {Nakamura}\ \emph {et~al.}(2016)\citenamefont
  {Nakamura}, \citenamefont {Ohya}, \citenamefont {Takahashi}, \citenamefont
  {Tsuruta}, \citenamefont {Sasaki}, \citenamefont {Uozumi}, \citenamefont
  {Norimatsu}, \citenamefont {Kitajima}, \citenamefont {Shikano},\ and\
  \citenamefont {Kayanuma}}]{Nakamura:16}%
  \BibitemOpen
  \bibfield  {author} {\bibinfo {author} {\bibfnamefont {K.~G.}\ \bibnamefont
  {Nakamura}}, \bibinfo {author} {\bibfnamefont {K.}~\bibnamefont {Ohya}},
  \bibinfo {author} {\bibfnamefont {H.}~\bibnamefont {Takahashi}}, \bibinfo
  {author} {\bibfnamefont {T.}~\bibnamefont {Tsuruta}}, \bibinfo {author}
  {\bibfnamefont {H.}~\bibnamefont {Sasaki}}, \bibinfo {author} {\bibfnamefont
  {S.}~\bibnamefont {Uozumi}}, \bibinfo {author} {\bibfnamefont
  {K.}~\bibnamefont {Norimatsu}}, \bibinfo {author} {\bibfnamefont
  {M.}~\bibnamefont {Kitajima}}, \bibinfo {author} {\bibfnamefont
  {Y.}~\bibnamefont {Shikano}}, \ and\ \bibinfo {author} {\bibfnamefont
  {Y.}~\bibnamefont {Kayanuma}},\ }\href {\doibase 10.1103/PhysRevB.94.024303}
  {\bibfield  {journal} {\bibinfo  {journal} {Phys. Rev. B}\ }\textbf {\bibinfo
  {volume} {94}},\ \bibinfo {pages} {024303} (\bibinfo {year}
  {2016})}\BibitemShut {NoStop}%
\bibitem [{\citenamefont {Sasaki}\ \emph {et~al.}(2018)\citenamefont {Sasaki},
  \citenamefont {Tanaka}, \citenamefont {Okano}, \citenamefont {Minami},
  \citenamefont {Kayanuma}, \citenamefont {Shikano},\ and\ \citenamefont
  {Nakamura}}]{Sasaki:18}%
  \BibitemOpen
  \bibfield  {author} {\bibinfo {author} {\bibfnamefont {H.}~\bibnamefont
  {Sasaki}}, \bibinfo {author} {\bibfnamefont {R.}~\bibnamefont {Tanaka}},
  \bibinfo {author} {\bibfnamefont {Y.}~\bibnamefont {Okano}}, \bibinfo
  {author} {\bibfnamefont {F.}~\bibnamefont {Minami}}, \bibinfo {author}
  {\bibfnamefont {Y.}~\bibnamefont {Kayanuma}}, \bibinfo {author}
  {\bibfnamefont {Y.}~\bibnamefont {Shikano}}, \ and\ \bibinfo {author}
  {\bibfnamefont {K.~G.}\ \bibnamefont {Nakamura}},\ }\href {\doibase
  10.1038/s41598-018-27734-1} {\bibfield  {journal} {\bibinfo  {journal} {Sci.
  Rep}\ }\textbf {\bibinfo {volume} {8}},\ \bibinfo {pages} {9609} (\bibinfo
  {year} {2018})}\BibitemShut {NoStop}%
\bibitem [{\citenamefont {Maehrlein}\ \emph {et~al.}(2017)\citenamefont
  {Maehrlein}, \citenamefont {Paarmann}, \citenamefont {Wolf},\ and\
  \citenamefont {Kampfrath}}]{Maehrlein:17}%
  \BibitemOpen
  \bibfield  {author} {\bibinfo {author} {\bibfnamefont {S.}~\bibnamefont
  {Maehrlein}}, \bibinfo {author} {\bibfnamefont {A.}~\bibnamefont {Paarmann}},
  \bibinfo {author} {\bibfnamefont {M.}~\bibnamefont {Wolf}}, \ and\ \bibinfo
  {author} {\bibfnamefont {T.}~\bibnamefont {Kampfrath}},\ }\href {\doibase
  10.1103/PhysRevLett.119.127402} {\bibfield  {journal} {\bibinfo  {journal}
  {Phys. Rev. Lett.}\ }\textbf {\bibinfo {volume} {119}},\ \bibinfo {pages}
  {127402} (\bibinfo {year} {2017})}\BibitemShut {NoStop}%
\bibitem [{\citenamefont {Shen}(1984)}]{Shen:01}%
  \BibitemOpen
  \bibfield  {author} {\bibinfo {author} {\bibfnamefont {Y.~R.}\ \bibnamefont
  {Shen}},\ }in\ \href@noop {} {\emph {\bibinfo {booktitle} {Principles of
  Nonlinear Optics}}}\ (\bibinfo  {publisher} {Wiley-Interscience},\ \bibinfo
  {year} {1984})\BibitemShut {NoStop}%
\bibitem [{\citenamefont {Zhang}\ \emph {et~al.}(2017)\citenamefont {Zhang},
  \citenamefont {Liu}, \citenamefont {Wu}, \citenamefont {Yi}, \citenamefont
  {Fang}, \citenamefont {Zhang}, \citenamefont {Zhong}, \citenamefont {Peng},
  \citenamefont {Liu},\ and\ \citenamefont {Song}}]{Zhang:16}%
  \BibitemOpen
  \bibfield  {author} {\bibinfo {author} {\bibfnamefont {B.}~\bibnamefont
  {Zhang}}, \bibinfo {author} {\bibfnamefont {S.}~\bibnamefont {Liu}}, \bibinfo
  {author} {\bibfnamefont {X.}~\bibnamefont {Wu}}, \bibinfo {author}
  {\bibfnamefont {T.}~\bibnamefont {Yi}}, \bibinfo {author} {\bibfnamefont
  {Y.}~\bibnamefont {Fang}}, \bibinfo {author} {\bibfnamefont {J.}~\bibnamefont
  {Zhang}}, \bibinfo {author} {\bibfnamefont {Q.}~\bibnamefont {Zhong}},
  \bibinfo {author} {\bibfnamefont {X.}~\bibnamefont {Peng}}, \bibinfo {author}
  {\bibfnamefont {X.}~\bibnamefont {Liu}}, \ and\ \bibinfo {author}
  {\bibfnamefont {Y.}~\bibnamefont {Song}},\ }\href {\doibase
  10.1016/j.ijleo.2016.11.107} {\bibfield  {journal} {\bibinfo  {journal}
  {Optik}\ }\textbf {\bibinfo {volume} {130}},\ \bibinfo {pages} {1073}
  (\bibinfo {year} {2017})}\BibitemShut {NoStop}%
\bibitem [{\citenamefont {Almeida}\ \emph {et~al.}(2017)\citenamefont
  {Almeida}, \citenamefont {Oncebay}, \citenamefont {Siqueira}, \citenamefont
  {Muniz}, \citenamefont {Boni},\ and\ \citenamefont {Mendonca}}]{Almeida:17}%
  \BibitemOpen
  \bibfield  {author} {\bibinfo {author} {\bibfnamefont {J.~M.~P.}\
  \bibnamefont {Almeida}}, \bibinfo {author} {\bibfnamefont {C.}~\bibnamefont
  {Oncebay}}, \bibinfo {author} {\bibfnamefont {J.~P.}\ \bibnamefont
  {Siqueira}}, \bibinfo {author} {\bibfnamefont {S.~R.}\ \bibnamefont {Muniz}},
  \bibinfo {author} {\bibfnamefont {L.~D.}\ \bibnamefont {Boni}}, \ and\
  \bibinfo {author} {\bibfnamefont {C.~R.}\ \bibnamefont {Mendonca}},\ }\href
  {\doibase 10.1038/s41598-017-14748-4} {\bibfield  {journal} {\bibinfo
  {journal} {Sci. Rep.}\ }\textbf {\bibinfo {volume} {7}},\ \bibinfo {pages}
  {14320} (\bibinfo {year} {2017})}\BibitemShut {NoStop}%
\bibitem [{\citenamefont {Dadap}\ \emph {et~al.}(1991)\citenamefont {Dadap},
  \citenamefont {Focht}, \citenamefont {Reitze},\ and\ \citenamefont
  {Downer}}]{Dadap:91}%
  \BibitemOpen
  \bibfield  {author} {\bibinfo {author} {\bibfnamefont {J.~I.}\ \bibnamefont
  {Dadap}}, \bibinfo {author} {\bibfnamefont {G.~B.}\ \bibnamefont {Focht}},
  \bibinfo {author} {\bibfnamefont {D.~H.}\ \bibnamefont {Reitze}}, \ and\
  \bibinfo {author} {\bibfnamefont {M.~C.}\ \bibnamefont {Downer}},\ }\href
  {\doibase 10.1364/OL.16.000499} {\bibfield  {journal} {\bibinfo  {journal}
  {Opt. Lett.}\ }\textbf {\bibinfo {volume} {16}},\ \bibinfo {pages} {499}
  (\bibinfo {year} {1991})}\BibitemShut {NoStop}%
\bibitem [{\citenamefont {Troj\'{a}nek}\ \emph {et~al.}(2010)\citenamefont
  {Troj\'{a}nek}, \citenamefont {Z\'{i}dek}, \citenamefont {Dzurn\'{a}k},
  \citenamefont {Koz\'{a}k},\ and\ \citenamefont {Mal\'{y}}}]{Trojanek:10}%
  \BibitemOpen
  \bibfield  {author} {\bibinfo {author} {\bibfnamefont {F.}~\bibnamefont
  {Troj\'{a}nek}}, \bibinfo {author} {\bibfnamefont {K.}~\bibnamefont
  {Z\'{i}dek}}, \bibinfo {author} {\bibfnamefont {B.}~\bibnamefont
  {Dzurn\'{a}k}}, \bibinfo {author} {\bibfnamefont {M.}~\bibnamefont
  {Koz\'{a}k}}, \ and\ \bibinfo {author} {\bibfnamefont {P.}~\bibnamefont
  {Mal\'{y}}},\ }\href {\doibase 10.1364/OE.18.001349} {\bibfield  {journal}
  {\bibinfo  {journal} {Opt. Exp.}\ }\textbf {\bibinfo {volume} {18}},\
  \bibinfo {pages} {1349} (\bibinfo {year} {2010})}\BibitemShut {NoStop}%
\bibitem [{\citenamefont {Sheik-Bahae}\ \emph {et~al.}(1990)\citenamefont
  {Sheik-Bahae}, \citenamefont {Said}, \citenamefont {Wei}, \citenamefont
  {Hagan},\ and\ \citenamefont {Stryland}}]{Bahae:90}%
  \BibitemOpen
  \bibfield  {author} {\bibinfo {author} {\bibfnamefont {M.}~\bibnamefont
  {Sheik-Bahae}}, \bibinfo {author} {\bibfnamefont {A.~A.}\ \bibnamefont
  {Said}}, \bibinfo {author} {\bibfnamefont {T.~H.}\ \bibnamefont {Wei}},
  \bibinfo {author} {\bibfnamefont {D.~J.}\ \bibnamefont {Hagan}}, \ and\
  \bibinfo {author} {\bibfnamefont {E.~W.~V.}\ \bibnamefont {Stryland}},\
  }\href {\doibase 10.1109/3.53394} {\bibfield  {journal} {\bibinfo  {journal}
  {IEEE J. Quantum Electron}\ }\textbf {\bibinfo {volume} {26}},\ \bibinfo
  {pages} {760} (\bibinfo {year} {1990})}\BibitemShut {NoStop}%
\bibitem [{\citenamefont {Meredith}(1982)}]{Meredith:82}%
  \BibitemOpen
  \bibfield  {author} {\bibinfo {author} {\bibfnamefont {G.~R.}\ \bibnamefont
  {Meredith}},\ }\href {\doibase 10.1063/1.443859} {\bibfield  {journal}
  {\bibinfo  {journal} {J. Chem. Phys}\ }\textbf {\bibinfo {volume} {77}},\
  \bibinfo {pages} {5863} (\bibinfo {year} {1982})}\BibitemShut {NoStop}%
\bibitem [{\citenamefont {Mondal}\ \emph {et~al.}(2018)\citenamefont {Mondal},
  \citenamefont {Saito}, \citenamefont {Aihara}, \citenamefont {Fons},
  \citenamefont {Kolobov}, \citenamefont {Tominaga}, \citenamefont {Murakami},\
  and\ \citenamefont {Hase}}]{Mondal:18}%
  \BibitemOpen
  \bibfield  {author} {\bibinfo {author} {\bibfnamefont {R.}~\bibnamefont
  {Mondal}}, \bibinfo {author} {\bibfnamefont {Y.}~\bibnamefont {Saito}},
  \bibinfo {author} {\bibfnamefont {Y.}~\bibnamefont {Aihara}}, \bibinfo
  {author} {\bibfnamefont {P.}~\bibnamefont {Fons}}, \bibinfo {author}
  {\bibfnamefont {A.~V.}\ \bibnamefont {Kolobov}}, \bibinfo {author}
  {\bibfnamefont {J.}~\bibnamefont {Tominaga}}, \bibinfo {author}
  {\bibfnamefont {S.}~\bibnamefont {Murakami}}, \ and\ \bibinfo {author}
  {\bibfnamefont {M.}~\bibnamefont {Hase}},\ }\href {\doibase
  10.1038/s41598-018-22196-x} {\bibfield  {journal} {\bibinfo  {journal} {Sci.
  Rep.}\ }\textbf {\bibinfo {volume} {8}},\ \bibinfo {pages} {3908} (\bibinfo
  {year} {2018})}\BibitemShut {NoStop}%
\bibitem [{\citenamefont {Maze}\ \emph {et~al.}(2011)\citenamefont {Maze},
  \citenamefont {Gali}, \citenamefont {Togan}, \citenamefont {Chu},
  \citenamefont {Trifonov}, \citenamefont {Kaxiras},\ and\ \citenamefont
  {Lukin}}]{Maze:11}%
  \BibitemOpen
  \bibfield  {author} {\bibinfo {author} {\bibfnamefont {J.~R.}\ \bibnamefont
  {Maze}}, \bibinfo {author} {\bibfnamefont {A.}~\bibnamefont {Gali}}, \bibinfo
  {author} {\bibfnamefont {E.}~\bibnamefont {Togan}}, \bibinfo {author}
  {\bibfnamefont {Y.}~\bibnamefont {Chu}}, \bibinfo {author} {\bibfnamefont
  {A.}~\bibnamefont {Trifonov}}, \bibinfo {author} {\bibfnamefont
  {E.}~\bibnamefont {Kaxiras}}, \ and\ \bibinfo {author} {\bibfnamefont
  {M.~D.}\ \bibnamefont {Lukin}},\ }\href {\doibase
  10.1088/1367-2630/13/2/025025} {\bibfield  {journal} {\bibinfo  {journal}
  {New J. Phys.}\ }\textbf {\bibinfo {volume} {13}},\ \bibinfo {pages} {025025}
  (\bibinfo {year} {2011})}\BibitemShut {NoStop}%
\bibitem [{\citenamefont {Hase}\ \emph {et~al.}(2012)\citenamefont {Hase},
  \citenamefont {Katsuragawa}, \citenamefont {Constantinescu},\ and\
  \citenamefont {Petek}}]{Hase:12}%
  \BibitemOpen
  \bibfield  {author} {\bibinfo {author} {\bibfnamefont {M.}~\bibnamefont
  {Hase}}, \bibinfo {author} {\bibfnamefont {M.}~\bibnamefont {Katsuragawa}},
  \bibinfo {author} {\bibfnamefont {A.~M.}\ \bibnamefont {Constantinescu}}, \
  and\ \bibinfo {author} {\bibfnamefont {H.}~\bibnamefont {Petek}},\ }\href
  {\doibase 10.1038/nphoton.2012.35} {\bibfield  {journal} {\bibinfo  {journal}
  {Nature Photon.}\ }\textbf {\bibinfo {volume} {6}},\ \bibinfo {pages} {243}
  (\bibinfo {year} {2012})}\BibitemShut {NoStop}%
\bibitem [{\citenamefont {Biersack}\ and\ \citenamefont
  {Ziegler}(1982)}]{Ziegler:85}%
  \BibitemOpen
  \bibfield  {author} {\bibinfo {author} {\bibfnamefont {J.~P.}\ \bibnamefont
  {Biersack}}\ and\ \bibinfo {author} {\bibfnamefont {J.~F.}\ \bibnamefont
  {Ziegler}},\ }in\ \href@noop {} {\emph {\bibinfo {booktitle} {Ion
  Implantation Techniques}}},\ \bibinfo {editor} {edited by\ \bibinfo {editor}
  {\bibfnamefont {H.}~\bibnamefont {Ryssel}}\ and\ \bibinfo {editor}
  {\bibfnamefont {H.}~\bibnamefont {Glawischnig}}}\ (\bibinfo  {publisher}
  {Springer},\ \bibinfo {year} {1982})\BibitemShut {NoStop}%
\bibitem [{\citenamefont {Kikuchi}\ \emph {et~al.}(2017)\citenamefont
  {Kikuchi}, \citenamefont {Prananto}, \citenamefont {Hayashi}, \citenamefont
  {Laraoui}, \citenamefont {Mizuochi}, \citenamefont {Hatano}, \citenamefont
  {Saitoh}, \citenamefont {Kim}, \citenamefont {Meriles},\ and\ \citenamefont
  {An}}]{Kikuchi:17}%
  \BibitemOpen
  \bibfield  {author} {\bibinfo {author} {\bibfnamefont {D.}~\bibnamefont
  {Kikuchi}}, \bibinfo {author} {\bibfnamefont {D.}~\bibnamefont {Prananto}},
  \bibinfo {author} {\bibfnamefont {K.}~\bibnamefont {Hayashi}}, \bibinfo
  {author} {\bibfnamefont {A.}~\bibnamefont {Laraoui}}, \bibinfo {author}
  {\bibfnamefont {N.}~\bibnamefont {Mizuochi}}, \bibinfo {author}
  {\bibfnamefont {M.}~\bibnamefont {Hatano}}, \bibinfo {author} {\bibfnamefont
  {E.}~\bibnamefont {Saitoh}}, \bibinfo {author} {\bibfnamefont
  {Y.}~\bibnamefont {Kim}}, \bibinfo {author} {\bibfnamefont {C.~A.}\
  \bibnamefont {Meriles}}, \ and\ \bibinfo {author} {\bibfnamefont
  {T.}~\bibnamefont {An}},\ }\href {\doibase 10.7567/APEX.10.103004} {\bibfield
   {journal} {\bibinfo  {journal} {Appl. Phys. Exp.}\ }\textbf {\bibinfo
  {volume} {10}},\ \bibinfo {pages} {103004} (\bibinfo {year}
  {2017})}\BibitemShut {NoStop}%
\bibitem [{\citenamefont {Pezzagna}\ \emph {et~al.}(2010)\citenamefont
  {Pezzagna}, \citenamefont {Naydenov}, \citenamefont {Jelezko}, \citenamefont
  {Wrachtrup},\ and\ \citenamefont {Meijer}}]{Pezzagna:10}%
  \BibitemOpen
  \bibfield  {author} {\bibinfo {author} {\bibfnamefont {S.}~\bibnamefont
  {Pezzagna}}, \bibinfo {author} {\bibfnamefont {B.}~\bibnamefont {Naydenov}},
  \bibinfo {author} {\bibfnamefont {F.}~\bibnamefont {Jelezko}}, \bibinfo
  {author} {\bibfnamefont {J.}~\bibnamefont {Wrachtrup}}, \ and\ \bibinfo
  {author} {\bibfnamefont {J.}~\bibnamefont {Meijer}},\ }\href {\doibase
  10.1088/1367-2630/12/6/065017} {\bibfield  {journal} {\bibinfo  {journal}
  {New J. Phys.}\ }\textbf {\bibinfo {volume} {12}},\ \bibinfo {pages} {065017}
  (\bibinfo {year} {2010})}\BibitemShut {NoStop}%
\bibitem [{\citenamefont {Yin}\ \emph {et~al.}(2000)\citenamefont {Yin},
  \citenamefont {Li}, \citenamefont {Tang},\ and\ \citenamefont {Ji}}]{Yin:00}%
  \BibitemOpen
  \bibfield  {author} {\bibinfo {author} {\bibfnamefont {M.}~\bibnamefont
  {Yin}}, \bibinfo {author} {\bibfnamefont {H.~P.}\ \bibnamefont {Li}},
  \bibinfo {author} {\bibfnamefont {S.~H.}\ \bibnamefont {Tang}}, \ and\
  \bibinfo {author} {\bibfnamefont {W.}~\bibnamefont {Ji}},\ }\href {\doibase
  10.1007/s003400050866} {\bibfield  {journal} {\bibinfo  {journal} {Appl.
  Phys. B}\ }\textbf {\bibinfo {volume} {70}},\ \bibinfo {pages} {587}
  (\bibinfo {year} {2000})}\BibitemShut {NoStop}%
\bibitem [{\citenamefont {Godfried}\ \emph {et~al.}(2015)\citenamefont
  {Godfried}, \citenamefont {Scarsbrook}, \citenamefont {Twitchen},
  \citenamefont {Houwman}, \citenamefont {Nelissen}, \citenamefont {Whitehead},
  \citenamefont {Hall},\ and\ \citenamefont {Martineau}}]{E6}%
  \BibitemOpen
  \bibfield  {author} {\bibinfo {author} {\bibfnamefont {H.~P.}\ \bibnamefont
  {Godfried}}, \bibinfo {author} {\bibfnamefont {G.~A.}\ \bibnamefont
  {Scarsbrook}}, \bibinfo {author} {\bibfnamefont {D.~J.}\ \bibnamefont
  {Twitchen}}, \bibinfo {author} {\bibfnamefont {E.~P.}\ \bibnamefont
  {Houwman}}, \bibinfo {author} {\bibfnamefont {W.~G.~M.}\ \bibnamefont
  {Nelissen}}, \bibinfo {author} {\bibfnamefont {A.~J.}\ \bibnamefont
  {Whitehead}}, \bibinfo {author} {\bibfnamefont {C.~E.}\ \bibnamefont {Hall}},
  \ and\ \bibinfo {author} {\bibfnamefont {P.~M.}\ \bibnamefont {Martineau}},\
  }in\ \href@noop {} {\emph {\bibinfo {booktitle} {US Patent}}}\ (\bibinfo
  {year} {2015})\ pp.\ \bibinfo {pages} {No.US 8,936,774 B2}\BibitemShut
  {NoStop}%
\bibitem [{\citenamefont {Liu}\ \emph {et~al.}(2017)\citenamefont {Liu},
  \citenamefont {Zhang}, \citenamefont {Zhong}, \citenamefont {Peng},\ and\
  \citenamefont {Liu}}]{Liu:17}%
  \BibitemOpen
  \bibfield  {author} {\bibinfo {author} {\bibfnamefont {X.}~\bibnamefont
  {Liu}}, \bibinfo {author} {\bibfnamefont {B.}~\bibnamefont {Zhang}}, \bibinfo
  {author} {\bibfnamefont {Q.}~\bibnamefont {Zhong}}, \bibinfo {author}
  {\bibfnamefont {X.}~\bibnamefont {Peng}}, \ and\ \bibinfo {author}
  {\bibfnamefont {S.}~\bibnamefont {Liu}},\ }\href {\doibase
  10.1088/1742-6596/867/1/012013} {\bibfield  {journal} {\bibinfo  {journal}
  {Journal of Physics: Conf. Series}\ }\textbf {\bibinfo {volume} {867}},\
  \bibinfo {pages} {012013} (\bibinfo {year} {2017})}\BibitemShut {NoStop}%
\bibitem [{\citenamefont {Sabbah}\ and\ \citenamefont
  {Riffe}(2002)}]{Sabbah:02}%
  \BibitemOpen
  \bibfield  {author} {\bibinfo {author} {\bibfnamefont {A.~J.}\ \bibnamefont
  {Sabbah}}\ and\ \bibinfo {author} {\bibfnamefont {D.~M.}\ \bibnamefont
  {Riffe}},\ }\href {\doibase 10.1103/PhysRevB.66.165217} {\bibfield  {journal}
  {\bibinfo  {journal} {Phys. Rev. B}\ }\textbf {\bibinfo {volume} {66}},\
  \bibinfo {pages} {165217} (\bibinfo {year} {2002})}\BibitemShut {NoStop}%
\bibitem [{\citenamefont {Zaitsev}(2001)}]{Zaitsev:01}%
  \BibitemOpen
  \bibfield  {author} {\bibinfo {author} {\bibfnamefont {A.~M.}\ \bibnamefont
  {Zaitsev}},\ }in\ \href {\doibase 10.1007/978-3-662-04548-0} {\emph {\bibinfo
  {booktitle} {Optical Properties of Diamond}}}\ (\bibinfo  {publisher}
  {Springer},\ \bibinfo {year} {2001})\BibitemShut {NoStop}%
\bibitem [{\citenamefont {Koz\'{a}k}\ \emph {et~al.}(2012)\citenamefont
  {Koz\'{a}k}, \citenamefont {Troj\'{a}nek}, \citenamefont {Dzur\v{n}\'{a}k},\
  and\ \citenamefont {Mal\'{y}}}]{Kozak:12}%
  \BibitemOpen
  \bibfield  {author} {\bibinfo {author} {\bibfnamefont {M.}~\bibnamefont
  {Koz\'{a}k}}, \bibinfo {author} {\bibfnamefont {F.}~\bibnamefont
  {Troj\'{a}nek}}, \bibinfo {author} {\bibfnamefont {B.}~\bibnamefont
  {Dzur\v{n}\'{a}k}}, \ and\ \bibinfo {author} {\bibfnamefont {P.}~\bibnamefont
  {Mal\'{y}}},\ }\href {\doibase 10.1364/JOSAB.29.001141} {\bibfield  {journal}
  {\bibinfo  {journal} {J. Opt. Soc. Am. B}\ }\textbf {\bibinfo {volume}
  {29}},\ \bibinfo {pages} {1141} (\bibinfo {year} {2012})}\BibitemShut
  {NoStop}%
\bibitem [{\citenamefont {Bourgeois}\ \emph {et~al.}(2015)\citenamefont
  {Bourgeois}, \citenamefont {Jarmola}, \citenamefont {Siyushev}, \citenamefont
  {Gulka}, \citenamefont {Hruby}, \citenamefont {Jelezko}, \citenamefont
  {Budker},\ and\ \citenamefont {Nesladek}}]{Bourgeois:15}%
  \BibitemOpen
  \bibfield  {author} {\bibinfo {author} {\bibfnamefont {E.}~\bibnamefont
  {Bourgeois}}, \bibinfo {author} {\bibfnamefont {A.}~\bibnamefont {Jarmola}},
  \bibinfo {author} {\bibfnamefont {P.}~\bibnamefont {Siyushev}}, \bibinfo
  {author} {\bibfnamefont {M.}~\bibnamefont {Gulka}}, \bibinfo {author}
  {\bibfnamefont {J.}~\bibnamefont {Hruby}}, \bibinfo {author} {\bibfnamefont
  {F.}~\bibnamefont {Jelezko}}, \bibinfo {author} {\bibfnamefont
  {D.}~\bibnamefont {Budker}}, \ and\ \bibinfo {author} {\bibfnamefont
  {M.}~\bibnamefont {Nesladek}},\ }\href {\doibase 10.1038/ncomms9577}
  {\bibfield  {journal} {\bibinfo  {journal} {Nat. Commun.}\ }\textbf {\bibinfo
  {volume} {6}},\ \bibinfo {pages} {8577} (\bibinfo {year} {2015})}\BibitemShut
  {NoStop}%
\bibitem [{\citenamefont {Wotherspoon}\ \emph {et~al.}(2003)\citenamefont
  {Wotherspoon}, \citenamefont {Steeds}, \citenamefont {Catmull},\ and\
  \citenamefont {Butler}}]{Wotherspoon:03}%
  \BibitemOpen
  \bibfield  {author} {\bibinfo {author} {\bibfnamefont {A.}~\bibnamefont
  {Wotherspoon}}, \bibinfo {author} {\bibfnamefont {J.}~\bibnamefont {Steeds}},
  \bibinfo {author} {\bibfnamefont {B.}~\bibnamefont {Catmull}}, \ and\
  \bibinfo {author} {\bibfnamefont {J.}~\bibnamefont {Butler}},\ }\href
  {\doibase 10.1016/S0925-9635(02)00229-7} {\bibfield  {journal} {\bibinfo
  {journal} {Diam. Relat. Mater.}\ }\textbf {\bibinfo {volume} {12}},\ \bibinfo
  {pages} {652} (\bibinfo {year} {2003})}\BibitemShut {NoStop}%
\bibitem [{\citenamefont {DeSalvo}\ \emph {et~al.}(1992)\citenamefont
  {DeSalvo}, \citenamefont {Hagan}, \citenamefont {Sheik-Bahae}, \citenamefont
  {Stegeman}, \citenamefont {Stryland},\ and\ \citenamefont
  {Vanherzeele}}]{DeSalvo:92}%
  \BibitemOpen
  \bibfield  {author} {\bibinfo {author} {\bibfnamefont {R.}~\bibnamefont
  {DeSalvo}}, \bibinfo {author} {\bibfnamefont {D.~J.}\ \bibnamefont {Hagan}},
  \bibinfo {author} {\bibfnamefont {M.}~\bibnamefont {Sheik-Bahae}}, \bibinfo
  {author} {\bibfnamefont {G.}~\bibnamefont {Stegeman}}, \bibinfo {author}
  {\bibfnamefont {E.~W.~V.}\ \bibnamefont {Stryland}}, \ and\ \bibinfo {author}
  {\bibfnamefont {H.}~\bibnamefont {Vanherzeele}},\ }\href {\doibase
  10.1364/OL.17.000028} {\bibfield  {journal} {\bibinfo  {journal} {Opt.
  Lett.}\ }\textbf {\bibinfo {volume} {17}},\ \bibinfo {pages} {28} (\bibinfo
  {year} {1992})}\BibitemShut {NoStop}%
\bibitem [{\citenamefont {Collins}(2002)}]{Collins:02}%
  \BibitemOpen
  \bibfield  {author} {\bibinfo {author} {\bibfnamefont {A.~T.}\ \bibnamefont
  {Collins}},\ }\href {\doibase 10.1088/0953-8984/14/14/307} {\bibfield
  {journal} {\bibinfo  {journal} {Journal of Phys.: Condens. Matter}\ }\textbf
  {\bibinfo {volume} {14}},\ \bibinfo {pages} {3743} (\bibinfo {year}
  {2002})}\BibitemShut {NoStop}%
\bibitem [{\citenamefont {Fuchs}\ \emph {et~al.}(2010)\citenamefont {Fuchs},
  \citenamefont {Dobrovitski}, \citenamefont {Toyli}, \citenamefont {Heremans},
  \citenamefont {Weis}, \citenamefont {Schenkel},\ and\ \citenamefont
  {Awschalom}}]{Fuchs:10}%
  \BibitemOpen
  \bibfield  {author} {\bibinfo {author} {\bibfnamefont {G.~D.}\ \bibnamefont
  {Fuchs}}, \bibinfo {author} {\bibfnamefont {V.~V.}\ \bibnamefont
  {Dobrovitski}}, \bibinfo {author} {\bibfnamefont {D.~M.}\ \bibnamefont
  {Toyli}}, \bibinfo {author} {\bibfnamefont {F.~J.}\ \bibnamefont {Heremans}},
  \bibinfo {author} {\bibfnamefont {C.~D.}\ \bibnamefont {Weis}}, \bibinfo
  {author} {\bibfnamefont {T.}~\bibnamefont {Schenkel}}, \ and\ \bibinfo
  {author} {\bibfnamefont {D.~D.}\ \bibnamefont {Awschalom}},\ }\href {\doibase
  10.1038/NPHYS1716} {\bibfield  {journal} {\bibinfo  {journal} {Nature Phys.}\
  }\textbf {\bibinfo {volume} {6}},\ \bibinfo {pages} {668} (\bibinfo {year}
  {2010})}\BibitemShut {NoStop}%
\bibitem [{\citenamefont {Subedi}\ \emph {et~al.}(2019)\citenamefont {Subedi},
  \citenamefont {Fedorov}, \citenamefont {Peppers}, \citenamefont {Martyshkin},
  \citenamefont {Mirov}, \citenamefont {Shao},\ and\ \citenamefont
  {Loncar}}]{Subedi:19}%
  \BibitemOpen
  \bibfield  {author} {\bibinfo {author} {\bibfnamefont {S.~D.}\ \bibnamefont
  {Subedi}}, \bibinfo {author} {\bibfnamefont {V.~V.}\ \bibnamefont {Fedorov}},
  \bibinfo {author} {\bibfnamefont {J.}~\bibnamefont {Peppers}}, \bibinfo
  {author} {\bibfnamefont {D.~V.}\ \bibnamefont {Martyshkin}}, \bibinfo
  {author} {\bibfnamefont {S.~B.}\ \bibnamefont {Mirov}}, \bibinfo {author}
  {\bibfnamefont {L.}~\bibnamefont {Shao}}, \ and\ \bibinfo {author}
  {\bibfnamefont {M.}~\bibnamefont {Loncar}},\ }\href {\doibase
  10.1364/OME.9.002076} {\bibfield  {journal} {\bibinfo  {journal} {Opt. Mater.
  Exp.}\ }\textbf {\bibinfo {volume} {9}},\ \bibinfo {pages} {2076} (\bibinfo
  {year} {2019})}\BibitemShut {NoStop}%
\bibitem [{\citenamefont {Aslam}\ \emph {et~al.}(2013)\citenamefont {Aslam},
  \citenamefont {Waldherr}, \citenamefont {Neumann}, \citenamefont {Jelezko},\
  and\ \citenamefont {Wrachtrup}}]{Aslam:13}%
  \BibitemOpen
  \bibfield  {author} {\bibinfo {author} {\bibfnamefont {N.}~\bibnamefont
  {Aslam}}, \bibinfo {author} {\bibfnamefont {G.}~\bibnamefont {Waldherr}},
  \bibinfo {author} {\bibfnamefont {P.}~\bibnamefont {Neumann}}, \bibinfo
  {author} {\bibfnamefont {F.}~\bibnamefont {Jelezko}}, \ and\ \bibinfo
  {author} {\bibfnamefont {J.}~\bibnamefont {Wrachtrup}},\ }\href {\doibase
  10.1088/1367-2630/15/1/013064} {\bibfield  {journal} {\bibinfo  {journal}
  {New J. Phys.}\ }\textbf {\bibinfo {volume} {15}},\ \bibinfo {pages} {013064}
  (\bibinfo {year} {2013})}\BibitemShut {NoStop}%
\bibitem [{\citenamefont {Dhomkar}\ \emph {et~al.}(2016)\citenamefont
  {Dhomkar}, \citenamefont {Henshaw}, \citenamefont {Jayakumar},\ and\
  \citenamefont {Meriles}}]{Dhomkar:16}%
  \BibitemOpen
  \bibfield  {author} {\bibinfo {author} {\bibfnamefont {S.}~\bibnamefont
  {Dhomkar}}, \bibinfo {author} {\bibfnamefont {J.}~\bibnamefont {Henshaw}},
  \bibinfo {author} {\bibfnamefont {H.}~\bibnamefont {Jayakumar}}, \ and\
  \bibinfo {author} {\bibfnamefont {C.~A.}\ \bibnamefont {Meriles}},\ }\href
  {\doibase 10.1126/sciadv.1600911} {\bibfield  {journal} {\bibinfo  {journal}
  {Sci. Adv.}\ }\textbf {\bibinfo {volume} {2}},\ \bibinfo {pages} {e1600911}
  (\bibinfo {year} {2016})}\BibitemShut {NoStop}%
\bibitem [{\citenamefont {Dhomkar}\ \emph {et~al.}(2018)\citenamefont
  {Dhomkar}, \citenamefont {Jayakumar}, \citenamefont {Zangara},\ and\
  \citenamefont {Meriles}}]{Dhomkar:18}%
  \BibitemOpen
  \bibfield  {author} {\bibinfo {author} {\bibfnamefont {S.}~\bibnamefont
  {Dhomkar}}, \bibinfo {author} {\bibfnamefont {H.}~\bibnamefont {Jayakumar}},
  \bibinfo {author} {\bibfnamefont {P.~R.}\ \bibnamefont {Zangara}}, \ and\
  \bibinfo {author} {\bibfnamefont {C.~A.}\ \bibnamefont {Meriles}},\ }\href
  {\doibase 10.1021/acs.nanolett.8b01739} {\bibfield  {journal} {\bibinfo
  {journal} {Nano Lett.}\ }\textbf {\bibinfo {volume} {18}},\ \bibinfo {pages}
  {4046} (\bibinfo {year} {2018})}\BibitemShut {NoStop}%
\bibitem [{\citenamefont {Babinec}\ \emph {et~al.}(2010)\citenamefont
  {Babinec}, \citenamefont {Hausmann}, \citenamefont {Khan}, \citenamefont
  {Zhang}, \citenamefont {Maze}, \citenamefont {Hemmer},\ and\ \citenamefont
  {Lon{\v c}ar}}]{Babinec:10}%
  \BibitemOpen
  \bibfield  {author} {\bibinfo {author} {\bibfnamefont {T.~M.}\ \bibnamefont
  {Babinec}}, \bibinfo {author} {\bibfnamefont {B.~J.~M.}\ \bibnamefont
  {Hausmann}}, \bibinfo {author} {\bibfnamefont {M.}~\bibnamefont {Khan}},
  \bibinfo {author} {\bibfnamefont {Y.}~\bibnamefont {Zhang}}, \bibinfo
  {author} {\bibfnamefont {J.~R.}\ \bibnamefont {Maze}}, \bibinfo {author}
  {\bibfnamefont {P.~R.}\ \bibnamefont {Hemmer}}, \ and\ \bibinfo {author}
  {\bibfnamefont {M.}~\bibnamefont {Lon{\v c}ar}},\ }\href {\doibase
  10.1038/nnano.2010.6} {\bibfield  {journal} {\bibinfo  {journal} {Nature
  Nanotech.}\ }\textbf {\bibinfo {volume} {5}},\ \bibinfo {pages} {195}
  (\bibinfo {year} {2010})}\BibitemShut {NoStop}%
\bibitem [{\citenamefont {Aharonovich}\ \emph {et~al.}(2011)\citenamefont
  {Aharonovich}, \citenamefont {Greentree},\ and\ \citenamefont
  {Prawer}}]{Aharonovich:11}%
  \BibitemOpen
  \bibfield  {author} {\bibinfo {author} {\bibfnamefont {I.}~\bibnamefont
  {Aharonovich}}, \bibinfo {author} {\bibfnamefont {A.~D.}\ \bibnamefont
  {Greentree}}, \ and\ \bibinfo {author} {\bibfnamefont {S.}~\bibnamefont
  {Prawer}},\ }\href {\doibase 10.1038/nphoton.2011.54} {\bibfield  {journal}
  {\bibinfo  {journal} {Nature Photon.}\ }\textbf {\bibinfo {volume} {5}},\
  \bibinfo {pages} {397} (\bibinfo {year} {2011})}\BibitemShut {NoStop}%
\bibitem [{\citenamefont {Hausmann}\ \emph {et~al.}(2014)\citenamefont
  {Hausmann}, \citenamefont {Bulu}, \citenamefont {Venkataraman}, \citenamefont
  {Deotare},\ and\ \citenamefont {Lon{\v c}ar}}]{Hausmann:14}%
  \BibitemOpen
  \bibfield  {author} {\bibinfo {author} {\bibfnamefont {B.~J.~M.}\
  \bibnamefont {Hausmann}}, \bibinfo {author} {\bibfnamefont {I.}~\bibnamefont
  {Bulu}}, \bibinfo {author} {\bibfnamefont {V.}~\bibnamefont {Venkataraman}},
  \bibinfo {author} {\bibfnamefont {P.}~\bibnamefont {Deotare}}, \ and\
  \bibinfo {author} {\bibfnamefont {M.}~\bibnamefont {Lon{\v c}ar}},\ }\href
  {\doibase 10.1038/nphoton.2014.72} {\bibfield  {journal} {\bibinfo  {journal}
  {Nature Photon.}\ }\textbf {\bibinfo {volume} {8}},\ \bibinfo {pages} {369}
  (\bibinfo {year} {2014})}\BibitemShut {NoStop}%
\end{thebibliography}%


\end{document}